\documentclass[%
 reprint,
 amsmath,amssymb,
 aps
]{revtex4-1}

\usepackage{graphicx, caption,subcaption}
\usepackage{float}
\usepackage{lipsum}
\usepackage{dcolumn}
\usepackage{bm}
\usepackage{xcolor} 

\begin{document}

\newcommand{\bra}[1]{\langle#1\rvert} 
\newcommand{\ket}[1]{\lvert#1\rangle} 

\title{Theory of momentum-resolved magnon electron energy loss spectra: The case of Yttrium Iron Garnet}

\author{Julio A. do Nascimento}
 \altaffiliation{julio.nascimento@york.ac.uk}
 \affiliation{School of Physics, Engineering and Technology, University of York, York YO10 5DD, UK}
 
\author{Phil J. Hasnip}%
\affiliation{School of Physics, Engineering and Technology, University of York, York YO10 5DD, UK}
\author{S. A. Cavill}%
\affiliation{School of Physics, Engineering and Technology, University of York, York YO10 5DD, UK}%

\author{Fabrizio Cossu}%

\affiliation{Department of Physics and Institute of Quantum Convergence and Technology,
Kangwon National University}

\author{Demie Kepaptsoglou}%
\affiliation{SuperSTEM, Sci-Tech Daresbury Campus}
\affiliation{School of Physics, Engineering and Technology, University of York, York YO10 5DD, UK}
\author{Quentin M. Ramasse}%
\affiliation{SuperSTEM, Sci-Tech Daresbury Campus}
\affiliation{School of Physics and Astronomy, University of Leeds}

\author{Adam Kerrigan}%
\affiliation{School of Physics, Engineering and Technology, University of York, York YO10 5DD, UK}
\affiliation{York-JEOL Nanocentre, University of York, York YO10 5BR, UK}
\author{Vlado K. Lazarov}%
\altaffiliation{vlado.lazarov@york.ac.uk}
\affiliation{School of Physics, Engineering and Technology, University of York, York YO10 5DD, UK}
\affiliation{York-JEOL Nanocentre, University of York, York YO10 5BR, UK}




\date{\today}

\begin{abstract}

We explore the inelastic spectra of electrons impinging on a magnetic system. The methodology here presented is intended to highlight the charge-dependent interaction of the electron beam in a STEM-EELS experiment, and the local vector potential generated by the magnetic lattice. This interaction shows an intensity similar to the purely spin interaction, which is taken to be functionally the same as in the inelastic neutron experiment. On the other hand, it shows a strong scattering vector dependence ($q^{-2}$), as well as a linear dependence with the probe's momentum. We also present the intensity dependence with the relative orientation between the probe wavevector and the local magnetic moments of the solid. We present YIG as a case study due to its high interest by the community.

\begin{description}
\item[Keywords] Spintronics, Magnonics, antiferromagnetic, yttrium iron garnet, YIG
\end{description}
\end{abstract}

\pacs{Valid PACS appear here}
\maketitle


\section{\label{intro}Introduction}

For quite some time, Moore's Law, especially the concern about its potential limits, has driven research into new computing approaches beyond traditional CMOS technology.
One approach that has attracted attention is to use the spin degree of freedom to substitute or integrate with the current electronic computation. To this objective, magnonics has been extensively studied, it encompasses the study of fundamental properties of magnons, which are quanta of the dynamic eigen-excitation of magnetically ordered materials in the form of spin-waves \cite{Mahmoud2020,Barman_2021}.

To systematically investigate the generation, manipulation, and identification of magnons, there is a requisite focus on improving the methodologies for both exciting and probing these phenomena. Magnons are commonly studied by inelastic neutron scattering (INS) techniques, time-resolved Kerr microscopy \cite{Kerr2017}, and Brillouin light scattering (BLS) \cite{BLS2015}. While these techniques probe the energy-momentum dispersion of magnons with high energy resolution, their spatial resolution is fundamentally limited to hundreds of nanometres.

Over the past decade, meV-level STEM-EELS has made significant strides, achieving atomic-level contrast\cite{Hage2019}, detecting spectral signatures of individual impurity atoms \cite{Hage2020}, and conducting spatial- and angle-resolved measurements on defects in crystalline materials \cite{Hoglund2022-qm}. 

The method's potential expansion into studying magnons is anticipated due to the overlapping energy range with vibrational modes in solid-state materials. Despite the weaker interaction of magnetic moments with the electron beam compared to the Coulomb potential, by up to 3 or 4 orders of magnitude, which makes their detection challenging,\cite{Loudon2012,Rusz2021} recent advancements in hybrid-pixel detectors, leading to a drastic improvement in the dynamic range and low background noise, with signals a mere 10$^{-7}$ of the full beam intensity readily detectable \cite{PLOTKINSWING2020}, and improved monochromator and spectrometer design, resulting in increased energy resolutions in particular at lower acceleration voltages (4.2meV at 30kV \cite{KRIVANEK201960}), offer enhanced sensitivity and signal detection, making the exploration of magnon excitation feasible in experimental settings.

Theoretical approaches for evaluating the EEL spectra have been developed focusing on spin-polarized probes interacting via an exchange-like interaction with the magnetic solid's local spins. At the core of the discussion is the understanding that from the point of view of the functional form of the spin-spin interaction, the neutron and electron scattering will be the same, the main difference that arises in the electron's case is its interaction with vector potential produced by the magnons in the system and, thus changing the probe's canonical momentum.

In our case we will explore the effects of the non-spin-polarized beam in the meV-level STEM-EELS apparatus, using YIG as a prototypical material. We will derive and compare the effects of the spin-based interaction and the charge-based interaction, and their momentum dependence.

\section{Methods}

The evaluation of the inelastic scattering of electrons by magnons requires the evaluation of the doubly differential cross-section. It evaluates the relative intensity of scattered particles into a solid angle $d\Omega$, with a wavevector in a small range around $\mathbf{k}_1$ given by $d\mathbf{k}_1$. Assuming N scatterers in the target, and a monochromatic beam with wavevector $\mathbf{k}_0$ in the z-direction with a particle current density $(J_0)_z$. This relative intensity can be written as \cite{Sturm1993,Stephen1984},


\begin{widetext}
\begin{equation}
\frac{d^{2} \sigma }{d\Omega d\mathbf{k}_{1}} =\frac{1}{N}\frac{N_{0} V\sum _{n_{0} ,n_{1},\sigma_{0}} P_{n_{0}}P_{\sigma_{0}}\mathbf{k}_{1}^{2} |\bra{n_{1}, \sigma_{1} ,\mathbf{k}_{1}} \hat{H}_{inter}\ket{n_{0}, \sigma_{0} ,\mathbf{k}_{0}} |^{2} \delta ( E_{n_{0}} +E_{0} -E_{n_{1}} -E_{1})}{( 2\pi )^{2} \hbar ( j_{0})_{z}}
    \label{eq:doubleDiff}
\end{equation}
\end{widetext}
where we are denoting a scattering process where the system undergoes a transition from state $n_0$ to $n_1$ with energies $E_0$ and $E_1$, respectively, simultaneously the scattered particle changes its state from momentum $k_0$ and spin $\sigma_0$ with energy $E_0$ to $k_1$, $\sigma_1$ with energy $E_1$. The interaction between the particle and the material is encapsulated by the interaction Hamiltonian $H_{inter}$. The current density of the particle beam along the z-direction is denoted by $( j_{0})_{z}$. Here, $P_{n_0}$ and $P_{\sigma_0}$ signifies the probability of the material to be in state $n_0$ and the beam to be in the spin state $\sigma_0$, before any scattering event. In this scenario, we begin by considering the initial state of the beam to be spin-unpolarized, which is formally represented by setting the spin polarization, $P_{\sigma_0}$, to one. Additionally, we assume that the creation of any active magnon mode is feasible, as magnons are bosons, and their creation is always possible. Therefore, we maintain $P_{n_0}=1$ throughout the analysis. Also, $N$, represents the number of statterers, $N_0$ is the number of particles in state $k_0$, and $V$ is the volume that defines the box normalization.

The choice of $H_{inter}$ is the central point of the discussion. In our case, we are interested in the interaction of an electron beam with the magnetic structure of the system. Disregarding the charge, the problem returns to a similar situation as to the INS, where the usual interaction is taken in terms of an interaction between the magnetic field generated by the intrinsic magnetic moments of the electrons and the orbital angular momentum of the system, and the probe's magnetic moment. This approach leads to a pair of terms, one exchange-like term and a term involving the spin-orbit interaction\cite{Stephen1984}. In the case of orbitally quenched systems, only the former term is taken into account. 
In addition to this spin-based interaction, the electron's charge must be considered when discussing EELS, specifically the alteration in the electron beam's canonical momentum, caused by the vector potential originating from the magnons in the magnetic solid. This interaction manifests exclusively when electrons serve as probes. For comprehensive coverage, we will study both spin-based and charge-based interactions, and study their similarities and differences.

\subsection{Spin-based interaction}

We start with the spin-based interaction. Here we will consider the interaction between the probe's magnetic moment and the magnetic field produced by the magnetic lattice.
We denote the spin of the electron beam in terms of the Pauli matrices $\bm{\hat{\sigma}}$, and assuming the g-factor $g_e=2$, we have the spin-based interaction to be given by,

\begin{equation}
    \hat{H}_{inter}^{SB} = -\mu_B\bm{\hat{\sigma}} \cdot \mathbf{H} 
\end{equation}
where $\mu_B$ is the Bohr magneton and $\mathbf{H}$ is the sample's magnetic field, which we will assume to be generated only by the set of local spins, ignoring the orbital contribution. Giving us, in terms of the spin operator $\mathbf{\hat{S}_j}$, where $j$ labels the different lattice positions of the spins in the solid,

\begin{equation}
     \hat{H}_{inter}^{SB} = 2\mu_{B}^{2}\bm{\hat{\sigma}} \cdot\left[\sum_{j}\nabla_{r} \times \left( \frac{\mathbf{\hat{S}}_j\times(\mathbf{r}-\mathbf{r}_j)}{|\mathbf{r}-\mathbf{r}_j|^3} \right) \right]
     \label{eq:interSB}
\end{equation}

Assuming that the scattering particle doesn't interact with the system before or after the scattering event, i.e., no multiple scattering events, we can write the total wave function as a product between a plane wave and the magnetic states,

\begin{equation}
    \ket{\mathbf{k}_i, \sigma_{i},n_i} \rightarrow \ket{\mathbf{k_i}, \sigma_{i}}\ket{n_i}
    \label{eq:BasisSeparated}
\end{equation}
for $i=0,1$. Here we have, $\ket{\mathbf{k}_i} =  \frac{1}{\sqrt{V}}e^{-i\mathbf{k}_{i} \cdot \mathbf{r}}$ being the state of the probing beam, while $\ket{n_i}$, represents the state of the solid, which for our purposes is only transiting between magnetic states, such that $\hat{H}_0\ket{n_i} = E_i\ket{n_i}$ with $\hat{H}_0$ being the Heisenberg Hamiltonian.

The procedure to derive the double-differential cross-section from the interaction given in Eq. \eqref{eq:interSB} follows the same steps for the neutron scattering case and can be found in the literature \cite{Lovesley,Squires2012}.

\begin{equation}
\frac{d^{2} \sigma }{d{\Omega}d\mathbf{k}_{1}} =\left(\frac{4\mu _{B}^{2} m_{e}}{\hbar ^{2}}\right)^{2}\frac{k_{1}}{k_{0}}\sum _{\alpha \beta }\mathcal{E}\left(\tilde{\mathbf{q}}\right)_{\alpha \beta }\mathcal{S}_{\alpha \beta } (\mathbf{q} ,\omega )
\label{eq:DoubleDiffFinal}
\end{equation}
with

\begin{equation}
\mathcal{E}\left(\tilde{\mathbf{q}}\right)_{\alpha \beta } =\begin{pmatrix}
1-\tilde{q}_{x}\tilde{q}_{x} & -\tilde{q}_{x}\tilde{q}_{y} & -\tilde{q}_{x}\tilde{q}_{z}\\
-\tilde{q}_{y}\tilde{q}_{x} & 1- \tilde{q}_{y}\tilde{q}_{y} & -\tilde{q}_{y}\tilde{q}_{z}\\
-\tilde{q}_{z}\tilde{q}_{x} & -\tilde{q}_{z}\tilde{q}_{y} & 1-\tilde{q}_{z}\tilde{q}_{z}
\end{pmatrix}
\end{equation}
where we have defined $\mathbf{q} = \mathbf{k}_1 - \mathbf{k}_0$, the scattering vector, its normalized form, $\mathbf{\tilde{q}}=\frac{\mathbf{q}}{q}$ and the spin-scattering function,

\begin{equation}
\begin{aligned}
    \mathcal{S}_{\alpha \beta }(\mathbf{q} ,\omega ) = \hspace{7.2cm} \\
    \frac{1}{2\pi N}\sum _{j,j'}F^{*}_{j}(\mathbf{q})F_{j'}(\mathbf{q})\int dte^{-i\omega t} e^{-i\mathbf{q} \cdot (\mathbf{r}_{j} -\mathbf{r}_{j'})} \langle S_{j}^{\alpha }( 0) S_{j'}^{\beta }( t) \rangle
\end{aligned}
    \label{eq:spinspinfunction_FF}
\end{equation}
where $F_{j}(\mathbf{q})$ and its conjugate $F^{*}_{j}(\mathbf{q})$ are the magnetic form factor. We define $\omega$ by, $\hbar \omega = E_{0}-E_{1}$ and $\langle S_{j}^{\alpha }( 0) S_{j'}^{\beta }( t) \rangle$ is the spin-spin correlation function.

\subsection{Charge-based interaction}

For the charge-based interaction, we have the result from \cite{Mendis2022},

\begin{equation}
H^{CB}_{inter} =-\frac{4\mu _{B}^{2}}{\hbar }\sum_j\mathbf{\hat{p}} \cdot \left(\mathbf{\hat{S}}_{j} \times \frac{(\mathbf{r} -\mathbf{r}_{j} )}{|\mathbf{r} -\mathbf{r}_{j} |^{3}}\right)\label{eq:ChargeInteraction}
\end{equation}
where we have $\mathbf{\hat{p}}$ the momentum operator, that acts on the incoming beam. This term comes from the vector potential term in the canonical momentum disregarding the weaker $\hat{\mathbf{A}}^2$ term, and the vector potential is taken as the total vector potential of the set of magnetic moments in sites $\mathbf{r}_{j}$.
Substituting the Eq. \eqref{eq:ChargeInteraction} into the Fermi Golden Rule of Eq. \eqref{eq:doubleDiff} we get,

\begin{widetext}
    \begin{equation}
        \bra{n_{1} ,\sigma _{1} ,\mathbf{k}_{1}} \hat{H}_{inter}^{CB}\ket{n_{0} ,\sigma _{0} ,\mathbf{k}_{0}} =\frac{4\mu _{B}^{2}}{\hbar V}  \bra{n_{1} ,\sigma _{1}} \int e^{-i\mathbf{k}_{1} \cdot (\mathbf{r} -\mathbf{r}_{j})}\left[\sum _{j} \mathbf{\hat{p}} \cdot \left(\frac{\bm{\hat{S}}_{j} \times \mathbf{r}}{|\textbf{r}|^{3}}\right)\right] e^{i\mathbf{k}_{0} \cdot (\mathbf{r} -\mathbf{r}_{j})}\ket{n_{0} ,\sigma _{0}}
    \end{equation}

    \begin{equation}
        \bra{n_{1} ,\sigma _{1} ,\mathbf{k}_{1}} \hat{H}_{inter}^{CB}\ket{n_{0} ,\sigma _{0} ,\mathbf{k}_{0}} = -\frac{4\mu _{B}^{2}}{V} i\bra{n_{1} ,\sigma _{1}} \int e^{-i(\mathbf{k}_{1} -\mathbf{k}_{0} )\cdot \mathbf{r}}\left[\sum _{j} e^{i(\mathbf{k}_{1} -\mathbf{k}_{0} )\cdot \mathbf{r}_{j}}( i\mathbf{k_{0}}) \cdot \left(\frac{\bm{\hat{S}}_{j} \times \mathbf{r}}{|\textbf{r}|^{3}}\right)\right] d\mathbf{r}\ket{n_{0} ,\sigma _{0}}
    \end{equation}
\end{widetext}
where we used $\mathbf{\hat{p}} = -i\hbar\nabla_r$, which is the evaluation of the momentum of the incoming beam. Taking the integral over $d\mathbf{r}$ is the same as taking the Fourier transform, where we used the known result from \cite{Fourier},

\begin{equation}
FT\left[\frac{\mathbf{r}}{|\mathbf{r} |^{3}}\right] =-\frac{4\pi i\mathbf{q}}{q^{2}}
\end{equation}
and again defining $\mathbf{q} = \mathbf{k}_{1} -\mathbf{k}_{0}$, the scattering vector, leading to,

\begin{equation}
\begin{aligned}
\bra{n_{1} ,\sigma _{1} ,\mathbf{k}_{1}} H_{inter}^{SB}\ket{n_{0} ,\sigma _{0} ,\mathbf{k}_{0}} =  \hspace{4.2cm}\\
-\frac{16\mu _{B}^{2} \pi i}{V} \bra{n_{1} ,\sigma _{1}}\int \sum _{j} e^{-i\mathbf{q} \cdot \mathbf{r}_{j}} \left[\mathbf{k_{0}} \cdot \left(\bm{\hat{S}}_{j} \times \frac{\mathbf{q}}{q^{2}}\right)\right]\ket{n_{0},\sigma _{0}}
\end{aligned}
\end{equation}
returning to Eq. \eqref{eq:doubleDiff}, taking $\displaystyle (j_{0} )_{z}$ as $\displaystyle \frac{N_{0}}{V}\frac{\hbar \mathbf{k}_{0}}{m_{e}}$, with $\displaystyle m_{e}$ being the mass of the electron, and $dE_{1} =\frac{\hbar ^{2} k_{1}}{m_e} dk_{1}$, we have,

\begin{widetext}
\begin{equation}
\frac{d^{2} \sigma }{d\omega d\mathbf{k}_{1}} =\left(\frac{8\mu _{B}^{2} m_{e}}{\hbar ^{2}}\right)^{2}\frac{1}{N}\frac{\mathbf{k}_{1}}{\mathbf{k}_{0}}\sum _{n_{0} ,n_{1}}  |\bra{n_{1},\sigma _{1}} \sum _{j} e^{-i\mathbf{q} \cdot \mathbf{r}_{j}} \left[\mathbf{k_{0}} \cdot \left(\bm{\hat{S}}_{j} \times \frac{\mathbf{q}}{|q|^{2}}\right)\right]\ket{n_{0},\sigma _{0}} |^{2} \delta (E_{n_{0}} +E_{0} -E_{n_{1}} -E_{1} )
\end{equation}
\end{widetext}
finally, assuming the incoming beam aligned with the z-direction $\displaystyle \mathbf{k_{0}} =k_{0z} \mathbf{\hat{z}}$, we finally have,
\begin{equation}
\frac{d^{2} \sigma }{d{\Omega}d\mathbf{k}_{1}} =\left(\frac{8\mu _{B}^{2} m_{e}}{\hbar ^{2}}\right)^{2} k_{1} k_{0z}\frac{1}{q^{2}}\sum _{\alpha \beta }\mathcal{E}\left(\tilde{\mathbf{q}}\right)_{\alpha \beta }\mathcal{S}_{\alpha \beta } (\mathbf{q} ,\omega )
\label{eq:ChargeDoubleDiff}
\end{equation}
with

\begin{equation}
\mathcal{E}\left(\tilde{\mathbf{q}}\right)_{\alpha \beta } =\begin{pmatrix}
\tilde{q}_{y}\tilde{q}_{y} & -\tilde{q}_{x}\tilde{q}_{y} & 0\\
-\tilde{q}_{y}\tilde{q}_{x} & \tilde{q}_{x}\tilde{q}_{x} & 0\\
0 & 0 & 0
\end{pmatrix}
\end{equation}
and Eq. \eqref{eq:spinspinfunction_FF}, where we used the identity \cite{Squires2012},

\begin{equation}
    \delta ( E_{n_{0}} +E_{0} -E_{n_{1}} -E_{1}) = \frac{1}{2\pi\hbar}\int e^{i(E_{n_{0}}-E_{n_{1}})t/\hbar}e^{i\omega t}dt
\end{equation}

Hence, the interaction can be thought to be the same as in a neutron scattering experiment, with the added effect of the charge-based interaction, which has a different momentum dependence. The calculation of the spin-scattering function $\displaystyle \mathcal{S}_{\alpha \beta } (\mathbf{q},\omega )$ can be done in a variety of ways, here we will use the approach described in \cite{Thz_Spectroscopy}, where the Holstein-Primakoff transformation is used and the bosonic creation and annihilation operators are projected on their diagonalised counterpart using the Bogoliubov transformation. Further more, the ’Kubler's trick' \cite{Kubler1988} is used, which allows for the calculation of the magnon modes with the freedom to align the local spins in an arbitrary orientation. In this trick the spin operator in the local reference frame ($\overline{\bm{\hat{S}}_{j}}$), is connected to the laboratory reference frame ($\bm{\hat{S}}_{j}$), by the unitary matrix $\mathcal{R}_{j}$ which is defined such that in the local reference frame, the spin always points in the z-direction, given by,

\begin{equation}
    \mathcal{R}_{j} =\begin{pmatrix}
cos\theta _{j} cos\phi _{j} & cos\theta_{j} sin\phi_{j} & -sin\theta_{j}\\
-sin\phi _{j} & cos\phi _{j} & 0\\
sin\theta _{j} cos\phi _{j} & sin\theta _{j} sin\phi _{j} & cos\theta _{j}
\end{pmatrix}
\label{eq:rotationMatrix}
\end{equation}
such that, $\overline{\bm{\hat{S}}_{j}} = \mathcal{R}_{j}\cdot\bm{\hat{S}}_{j}$, where $\theta_j$ and $\phi_j$ are the polar and azimuthal angles, respectively, of the spins in the laboratory reference frame.

One point to note is the coupling constant, in this case, $\left(\frac{8\mu _{B}^{2} m_{e}}{\hbar ^{2}}\right)^{2} = 0.3176$ Barn. Note that similarly to the case of inelastic scattering of electrons by phonons, the magnon case exhibits an overall dependence of $q^{-2}$. Hence, corroborating with the discussion in \cite{Nicholls2019} we expect the signal to be strongest in the first Brillouin zone, in contrast with the cases when neutrons or photons are used, where the data has a stronger signal for larger $q$. In addition, the linear $k_0$ dependence in the charge-based interaction, points to the possibility to enhance the effect with higher acceleration voltages in the STEM-EELS experiment.

In the next section, we will compare the two interactions and compare the results in inelastic neutron scattering (INS) which is a well-regarded probing method for momentum-resolved analysis of quasi-particle dispersion relations \cite{HudsonINS}. The spin-based interaction has the same functional form as the theory of INS, differing in the coupling constant.

For our analysis, we will use YIG as a prototypical material for study, due to the high interest of the community in its Thz magnons capability and high free-path length.

For the calculation of the underlying magnon dispersion, we will use the exchange parameters proposed from fitting inelastic neutron scattering in \cite{Princep2017}.

\begin{figure}[h!]
\centering
    \includegraphics[scale=0.18]{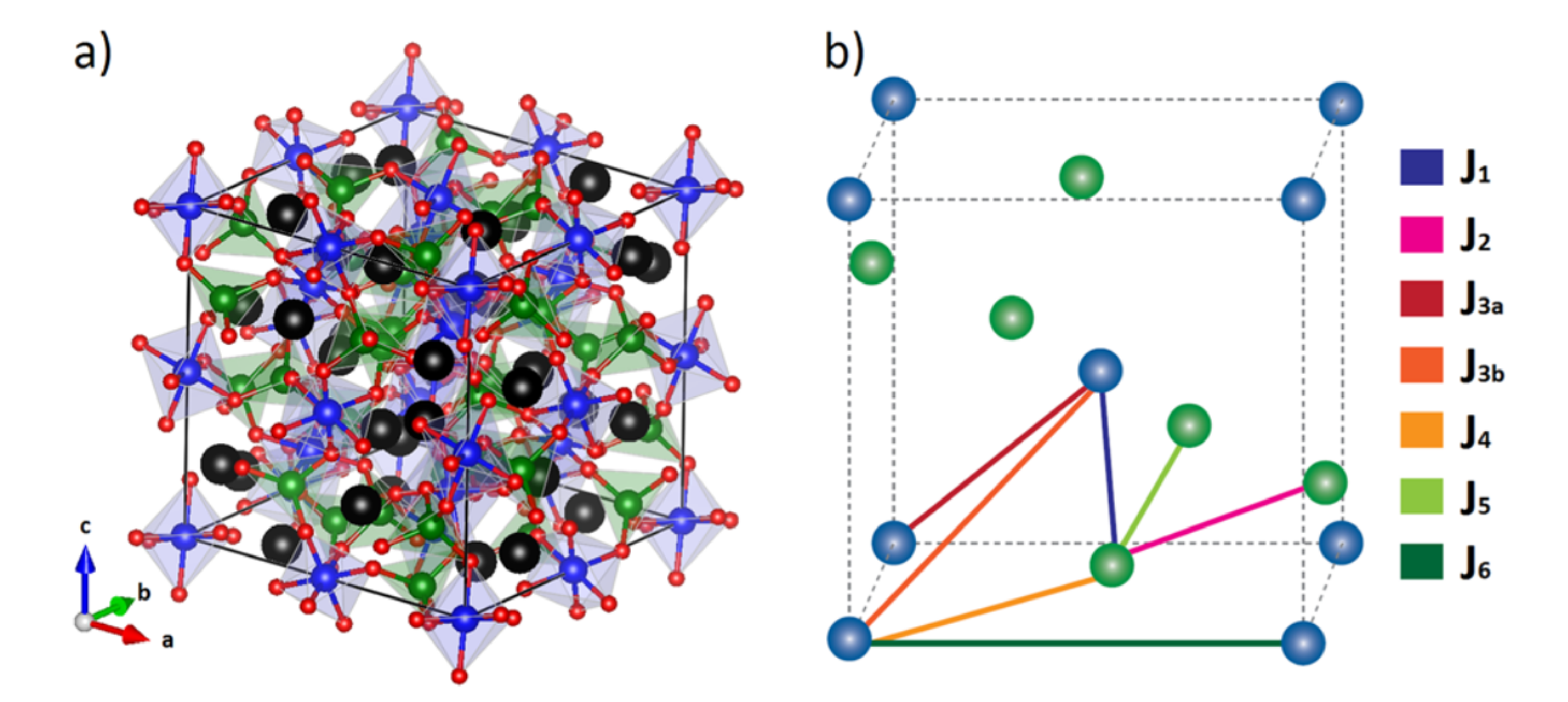}
    \caption{Crystal structure and magnetic exchange paths in Yttrium Iron Garnet (YIG). The conventional unit cell of YIG is represented, with the majority tetrahedral sites marked in green and the minority octahedral sites in blue. Yttrium is depicted as black spheres, and oxygen as red spheres. The first octant of the YIG unit cell is shown, highlighting the two distinct Fe3+ sites: tetrahedral sites in green and octahedral sites in blue \cite{Princep2017}.}
    \label{fig:enter-label}
\end{figure}

Note that, for the third nearest neighbours, there are two possibilities of exchange parameters $J_{3a}$ and $J_{3b}$. These two exchange paths are dissimilar and can be distinguished due to the symmetry of the crystal when rotated about the bond vector. The $J_{3a}$ exchange path exhibits a 2-fold symmetry, while the $J_{3b}$ exchange path obey the higher D3 symmetry point group \cite{Princep2017}.

For the magnetic form factor, we used the values for Fe$^{+3}$ given in \cite{Brown}.

We point out that the coupling constant for EELS is in the same order of magnitude as the neutron's. The main experimental difference lies in the flux of particles in the two experiments. The neutron scattering experiment has a typical flux of $10^{14}$ neutrons cm$^{-2}$s$^{-1}$ \cite{NeutronFlux}, which is $10^5$ times lower than the typical $10^{19}$ electrons cm$^{-2}$s$^{-1}$\cite{ElectronFlux} in an electron microscope, leading to a lower exposition time required by the EELS experiment compared to the INS.

\section{Results}\label{Results}

The distinctions between the interaction forms stem from their dependence on three factors: the variable $q$, the variation in intensity relative to the angle between the beam's orientation and that of the local magnetic moments, and primarily, the dependence on the beam's initial momentum. The charge-based interaction is linear with the beam momentum $k_0$ while the spin-based follows a $k^{-1}_0$ relation, hence the effect will be dominant for higher acceleration voltages.

\begin{figure*}[h!]
    \centering 
\begin{subfigure}{0.316455696\textwidth}
    \raggedright
    (a) 
    \includegraphics[width=\linewidth]{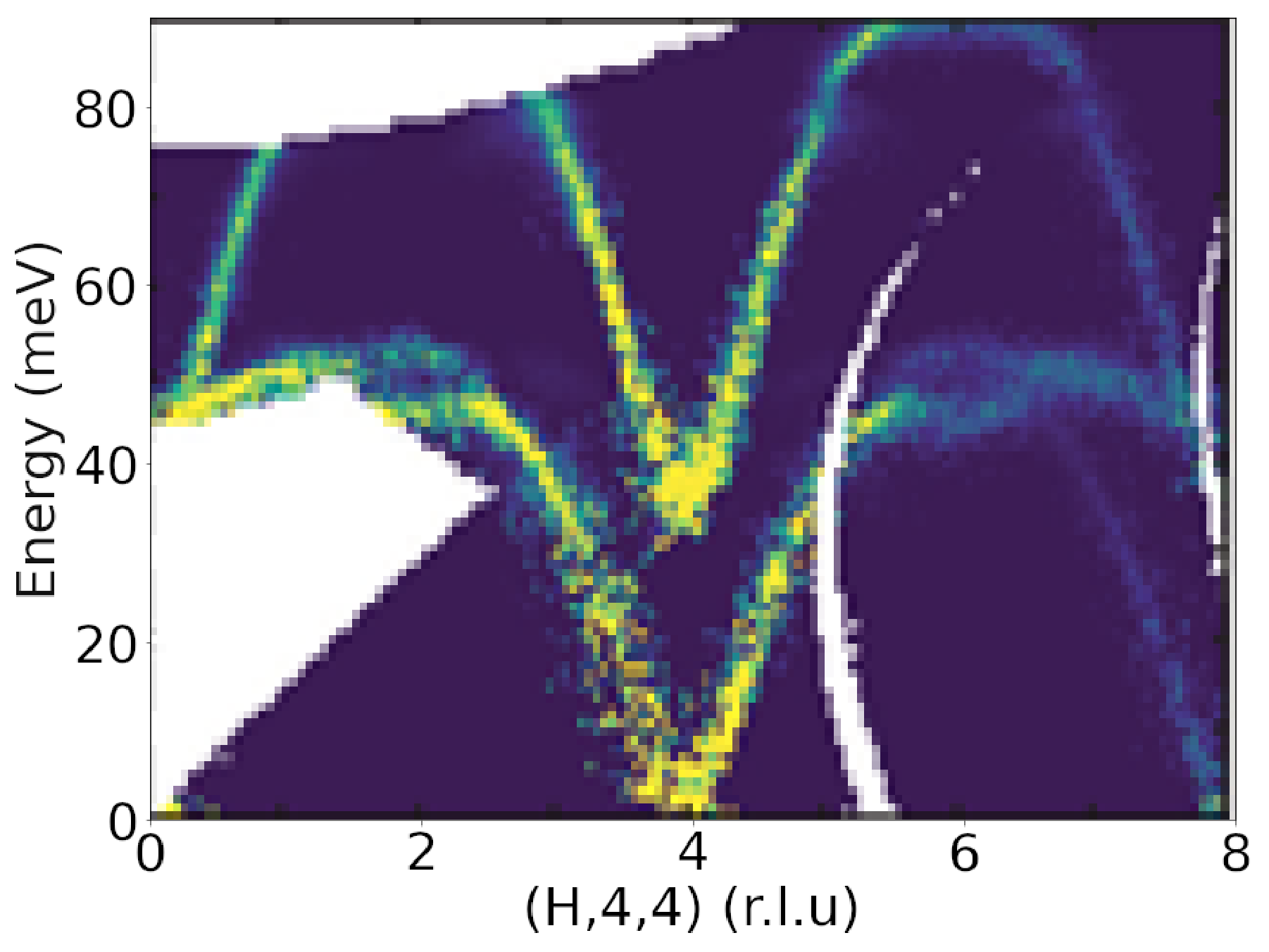}

  \label{fig:1}
\end{subfigure}\hfil 
\begin{subfigure}{0.341772152\textwidth}
    \raggedright
    (b) Spin-based
    \includegraphics[width=\linewidth]{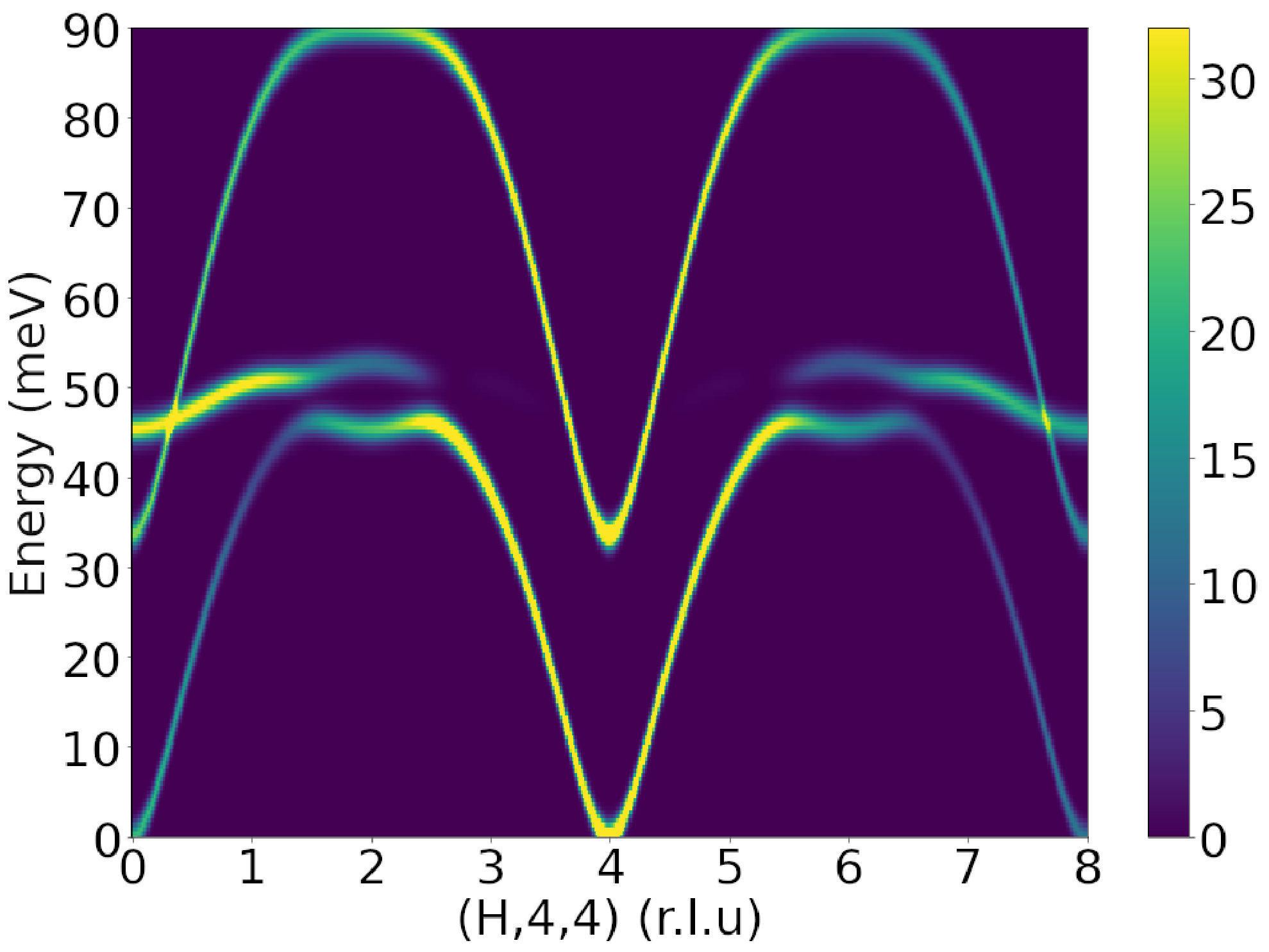}
    
  \label{fig:2}
\end{subfigure}\hfil 
\begin{subfigure}{0.341772152\textwidth}
    \raggedright
    (c) Charge-based
  \includegraphics[width=\linewidth]{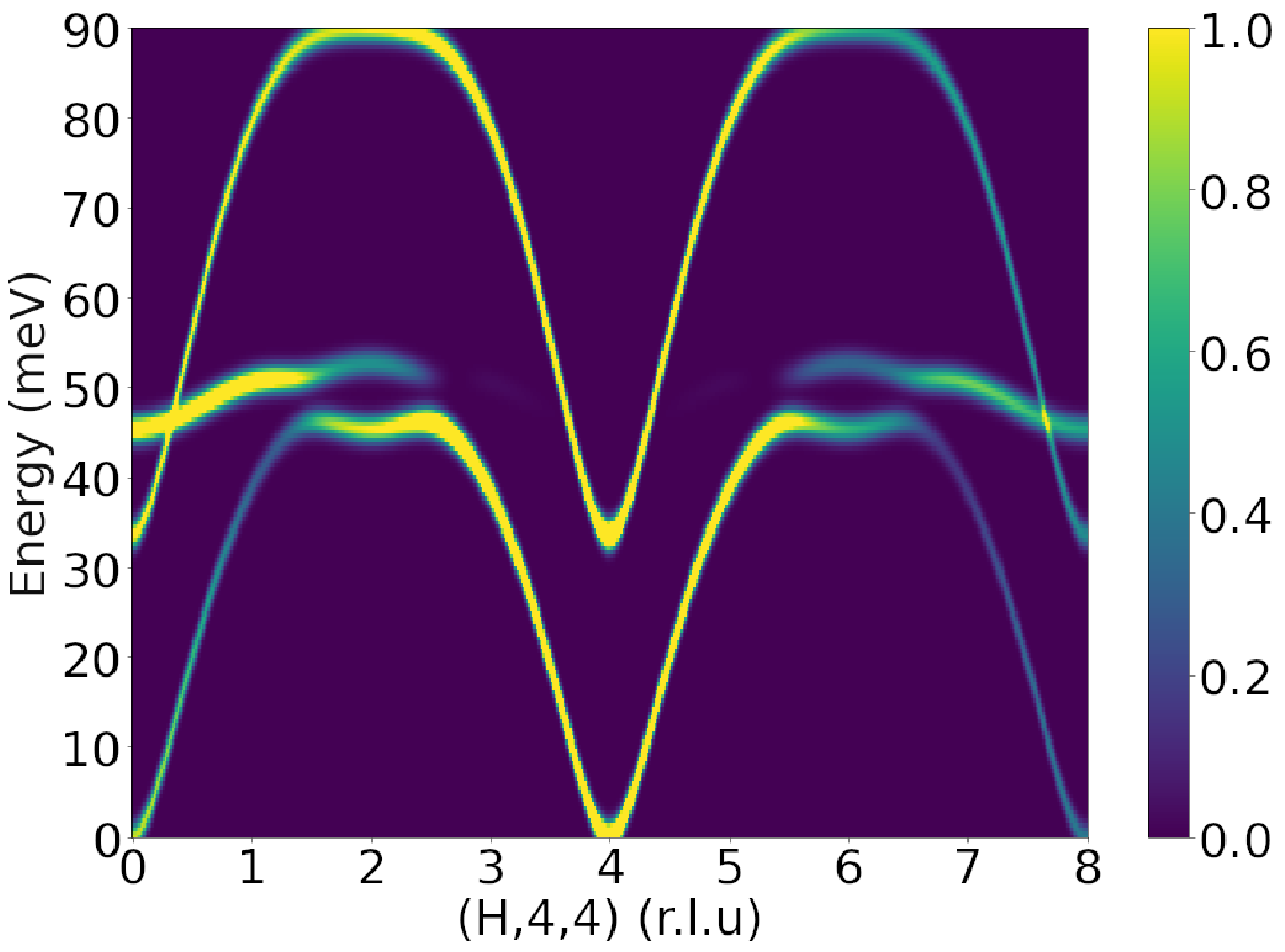}

  \label{fig:3}
\end{subfigure}
\caption{Inelastic scattering by magnons: a) Experimental inelastic neutron scattering \cite{Princep2017}, b) theoretical evaluation of spin-based EELS, c) charge-based EELS. All the calculations were performed for a relative angle between the probe's wave vector and the Nèel vector $\theta=0$.}
\label{fig:ComparissonINSEELS}
\end{figure*}

In figure \ref{fig:ComparissonINSEELS} we compare the experimentally acquired INS given in \cite{Princep2017} with the calculated INS using Eq. \eqref{eq:DoubleDiffFinal} and the calculated charge-only EELS using Eq. \eqref{eq:ChargeDoubleDiff}. 

Taking the definition of the spin orientation given in \ref{eq:rotationMatrix}, we kept the orientation of the magnetic moments aligned with oriented parallel to the axis of $\theta=0$ and $\phi=0$, while the electron/neutron beam is kept along the z-axis. We can see the similarities between the experiment and the calculated spectra for the EELS, particularly the same modes are active. For this comparison, we assumed $k_0=1$ for simplicity, and the intensity difference between the spin-based and change-based interaction comes mainly due to the $q$ dependence. Under the discussion made before, both the intensities given in the figures \ref{fig:ComparissonINSEELS}-b and  \ref{fig:ComparissonINSEELS}-c coexist in the EELS spectra. 

Taking the definition of the spin orientation given in \ref{eq:rotationMatrix}, we will keep $\phi=0$ and change the value of $\theta$. In figure \ref{fig:ComparissonINSEELS} 

In figure \ref{fig:YIG_orientation} we kept $\phi=0$ and changed the value of $\theta$, to probe the effect of changing the relative angle between the probe wavevector and the local magnetic moments. We see a strong dependence on the magnetic moments orientation in the charge interaction case. Here we see that the proposed interaction shows a dependence on the intensity as we change the polar angle $\theta$ of the orientation magnetic moments axis of orientation. In the path $\Gamma-H$ the intensity varies with a cosine relation with the angle between the spin and the beam angles.

\begin{figure*}[h!]
    \centering 
\begin{subfigure}{0.333333\textwidth}
    \raggedright
    (a) $\theta = 0$
  \includegraphics[width=\linewidth]{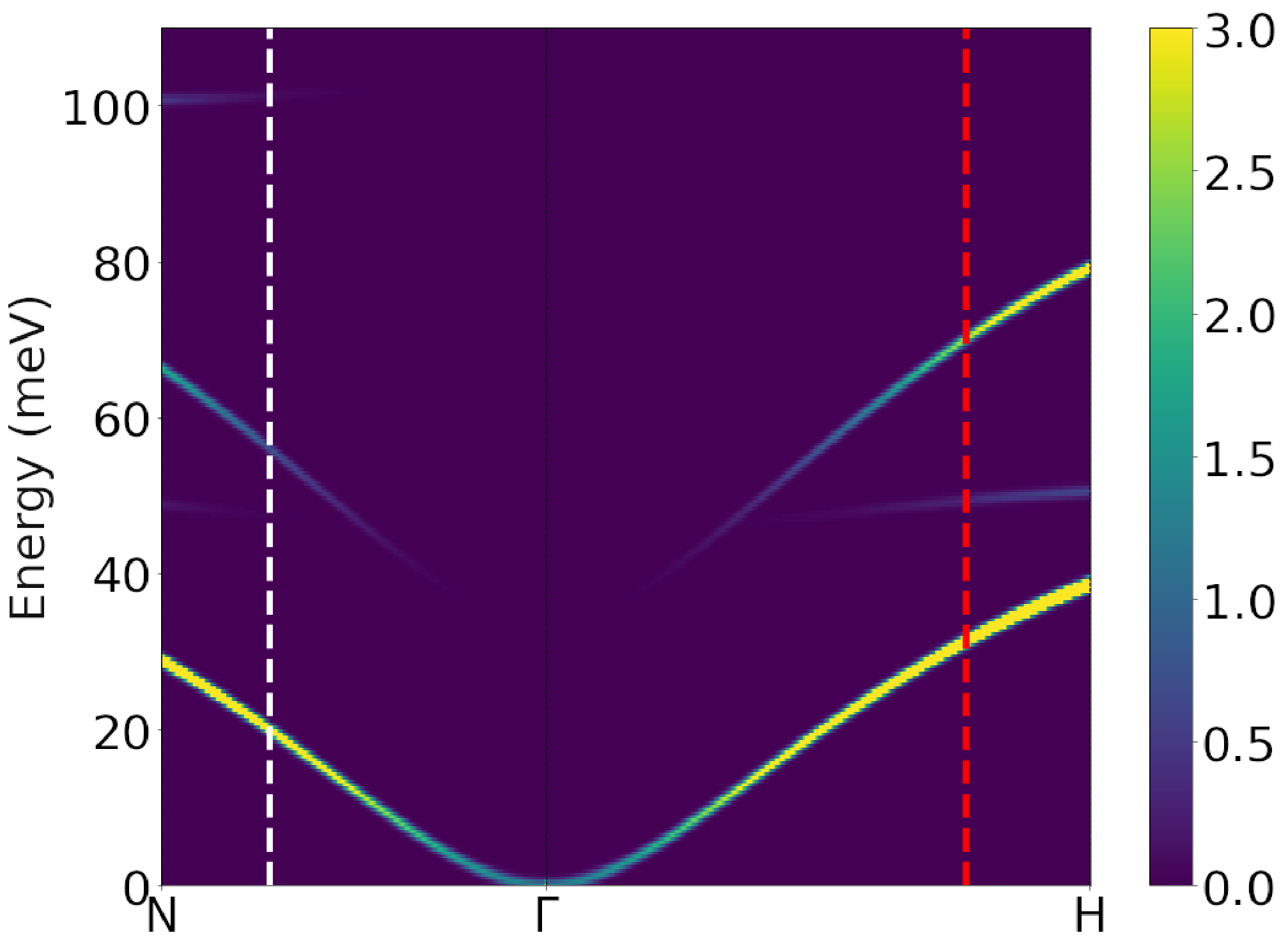}

  \label{fig:NiO_1}
\end{subfigure}\hfil 
\begin{subfigure}{0.333333\textwidth}
    \raggedright
    (b) $\theta = 3\pi/10$
  \includegraphics[width=\linewidth]{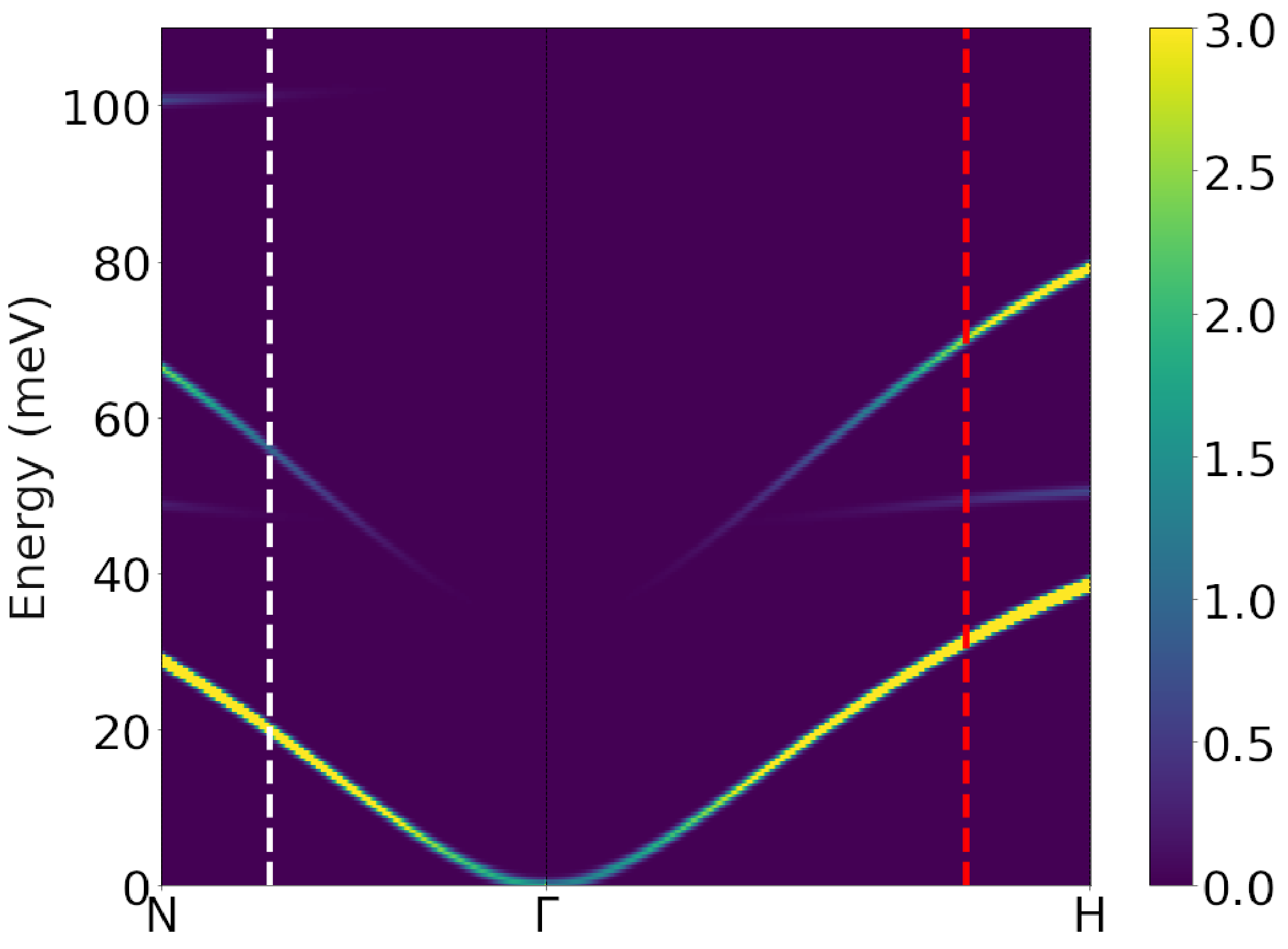}

  \label{fig:NiO_2}
\end{subfigure}\hfil 
\begin{subfigure}{0.333333\textwidth}
    \raggedright
    (c) $\theta = \pi/2$
  \includegraphics[width=\linewidth]{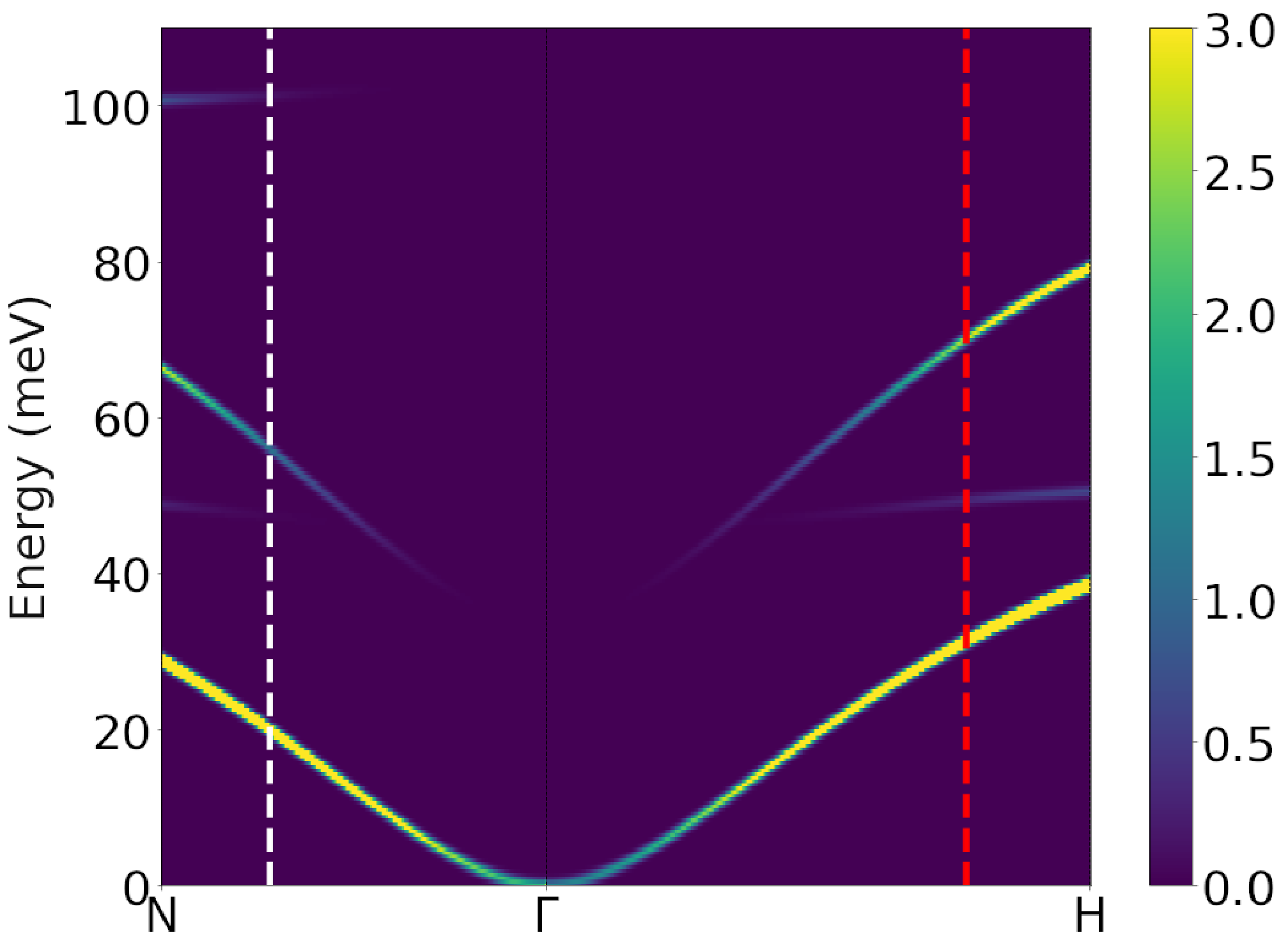}

  \label{fig:NiO_3}
\end{subfigure}

\medskip
\begin{subfigure}{0.3333336\textwidth}
    \raggedright
    (d) $\theta = 7\pi/10$
  \includegraphics[width=\linewidth]{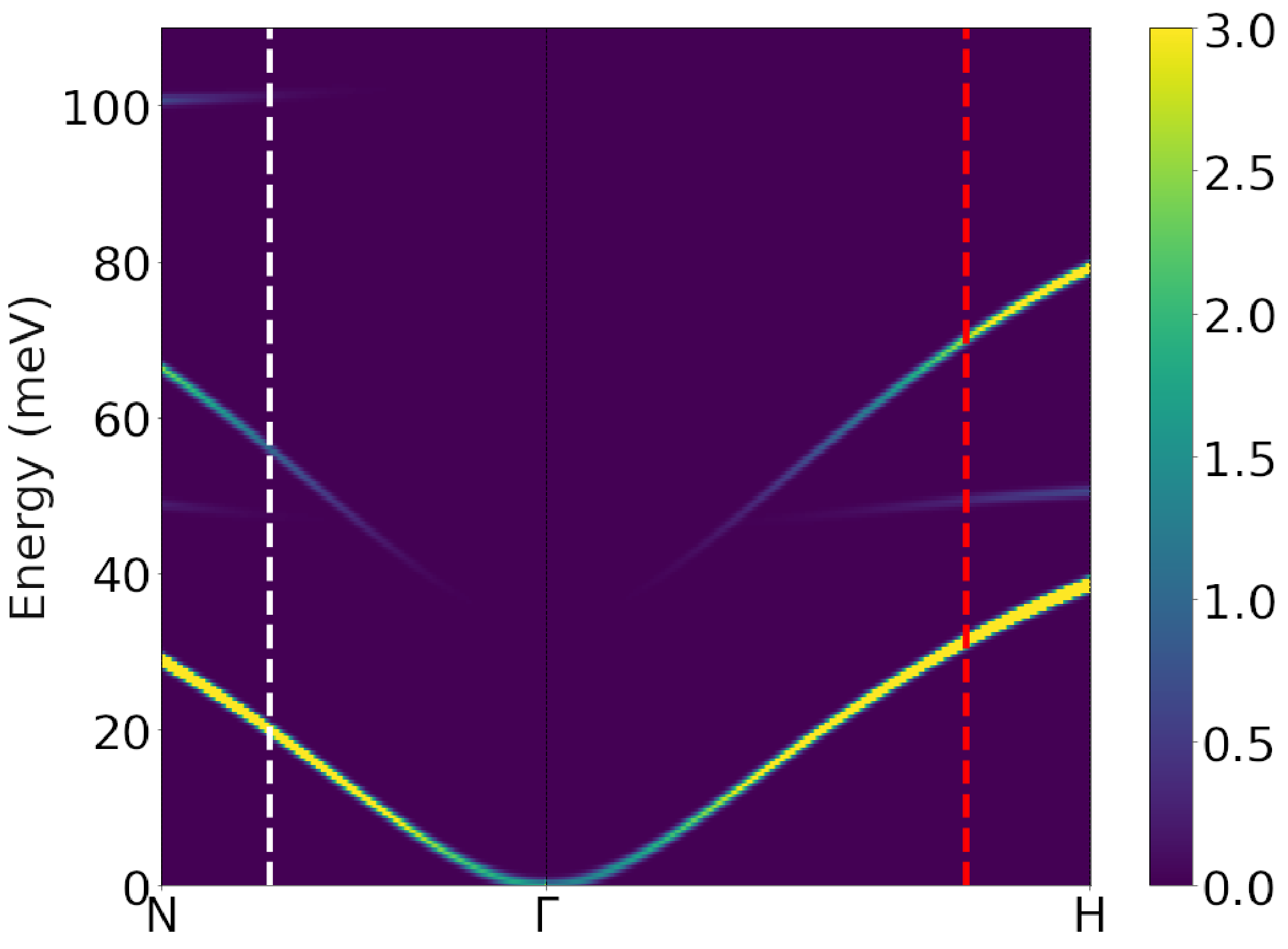}

  \label{fig:NiO_4}
\end{subfigure}\hfil 
\begin{subfigure}{0.333333\textwidth}
    \raggedright
    (e) 
  \includegraphics[width=\linewidth]{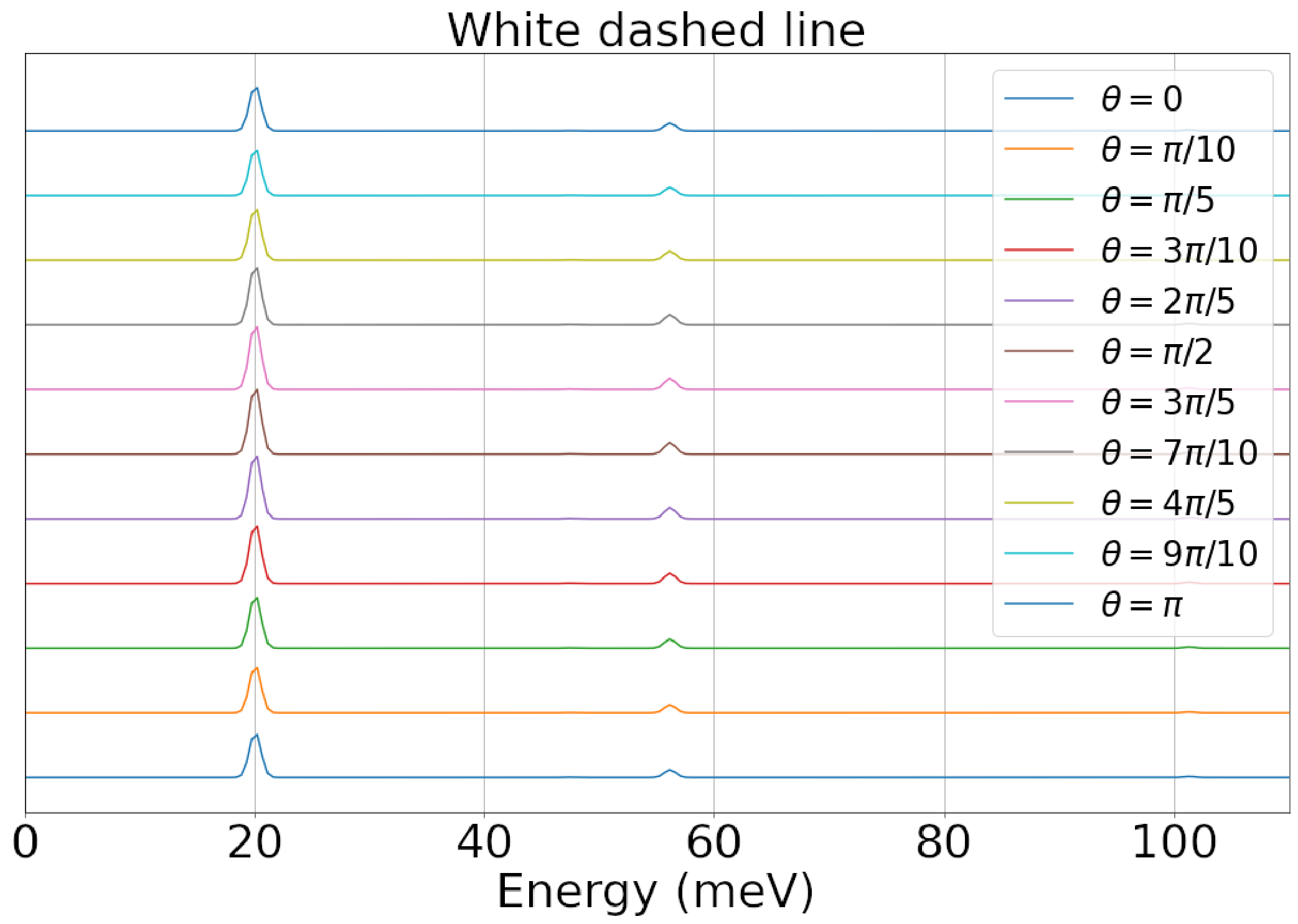}

  \label{fig:NiO_5}
\end{subfigure}\hfil 
\begin{subfigure}{0.333333\textwidth}
    \raggedright
    (f)
  \includegraphics[width=\linewidth]{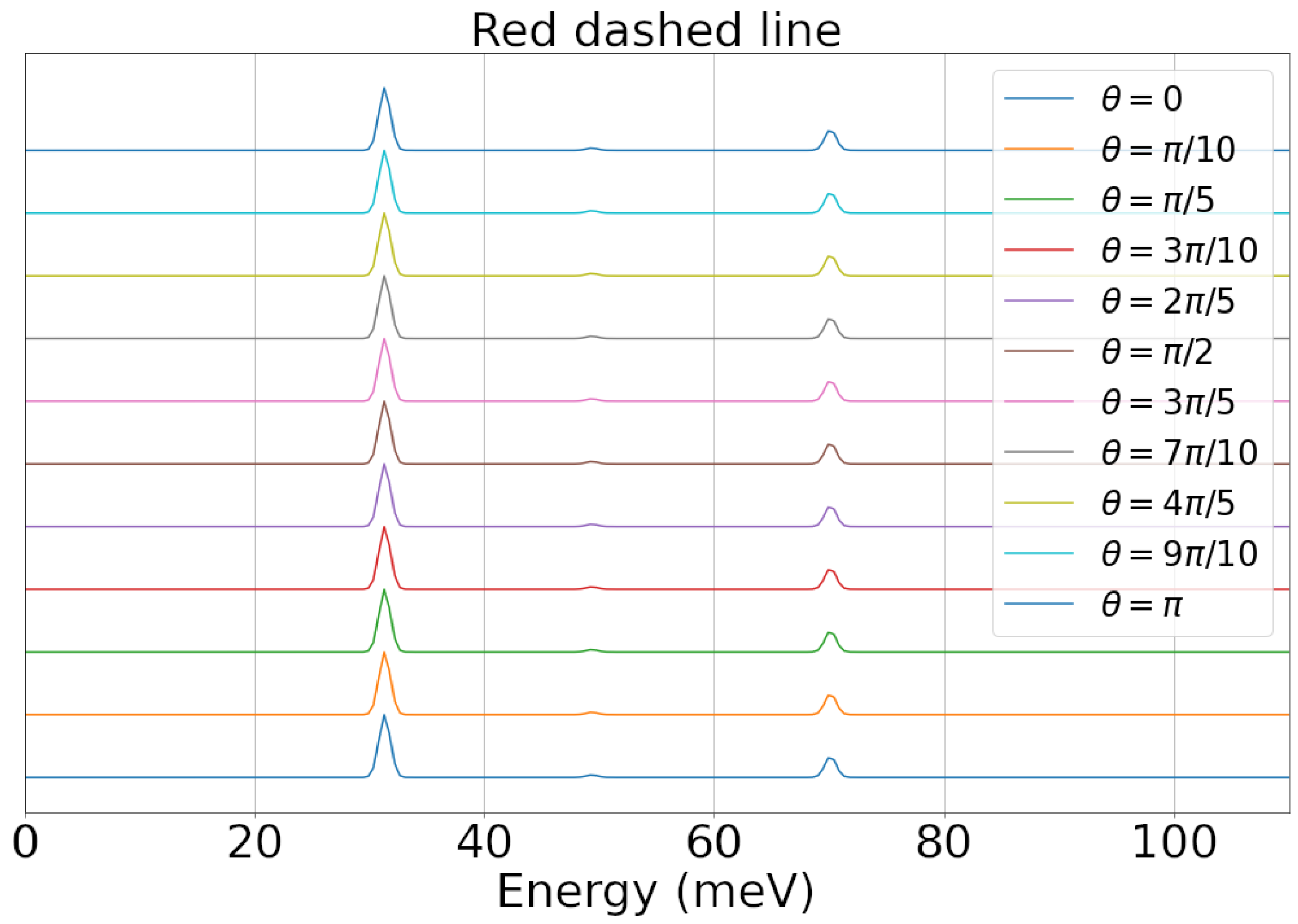}

  \label{fig:NiO_6}
\end{subfigure}
\caption{a-d) Spin-related EELS, for varying relative angles $\theta$ between the probe's wave vector and the Nèel vector. e-f) Angle-dependent intensity for the points in momentum space represented by the white and red dashed lines, showing a weak angle dependence on the point represented by the red and white dashed lines.}
\label{fig:YIG_orientation}
\end{figure*}

\begin{figure*}[h!]
    \centering 
\begin{subfigure}{0.333333\textwidth}
    \raggedright
    (a) $\theta = 0$
  \includegraphics[width=\linewidth]{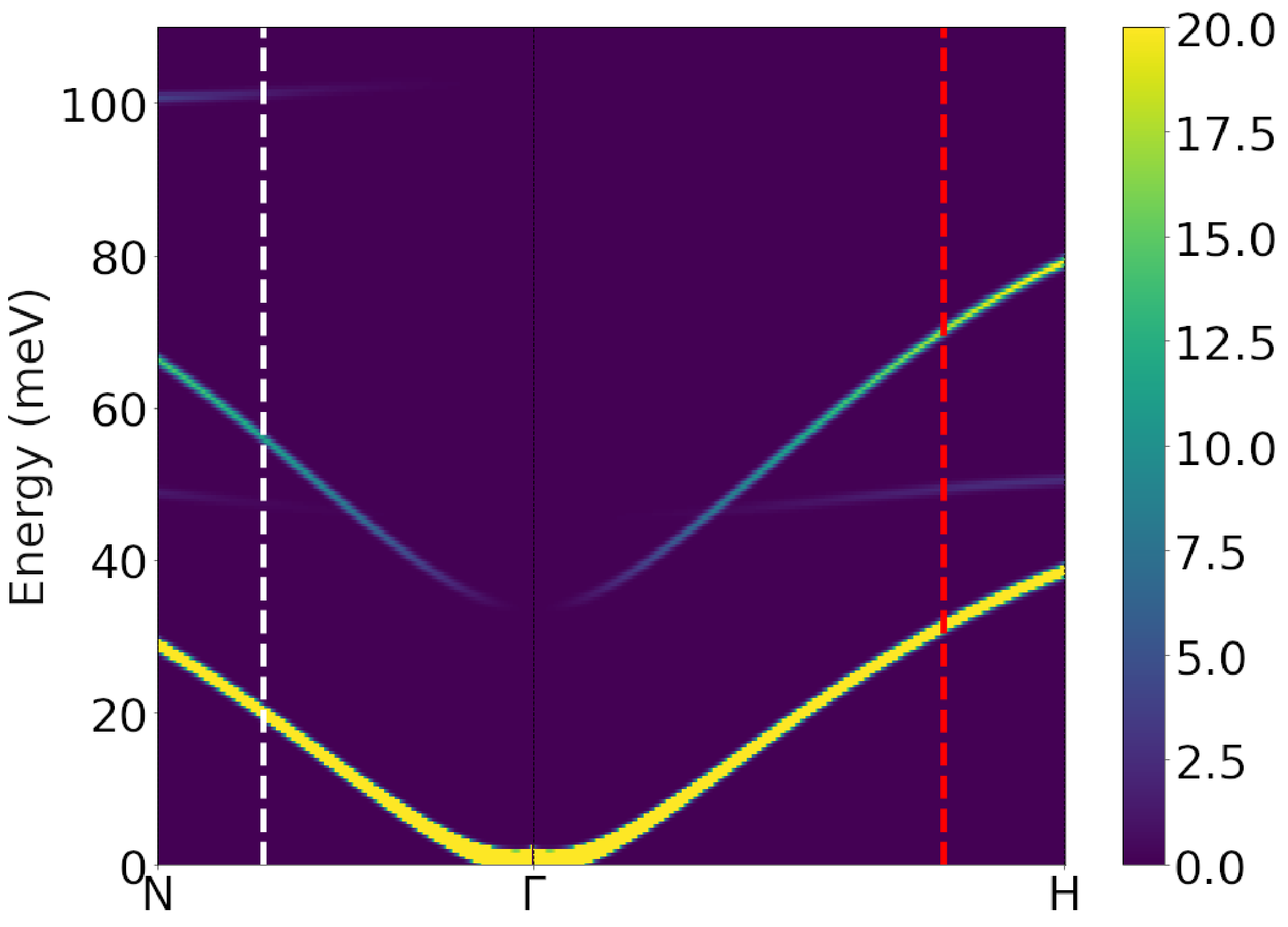}

  \label{fig:NiO_1}
\end{subfigure}\hfil 
\begin{subfigure}{0.333333\textwidth}
    \raggedright
    (b) $\theta = 3\pi/10$
  \includegraphics[width=\linewidth]{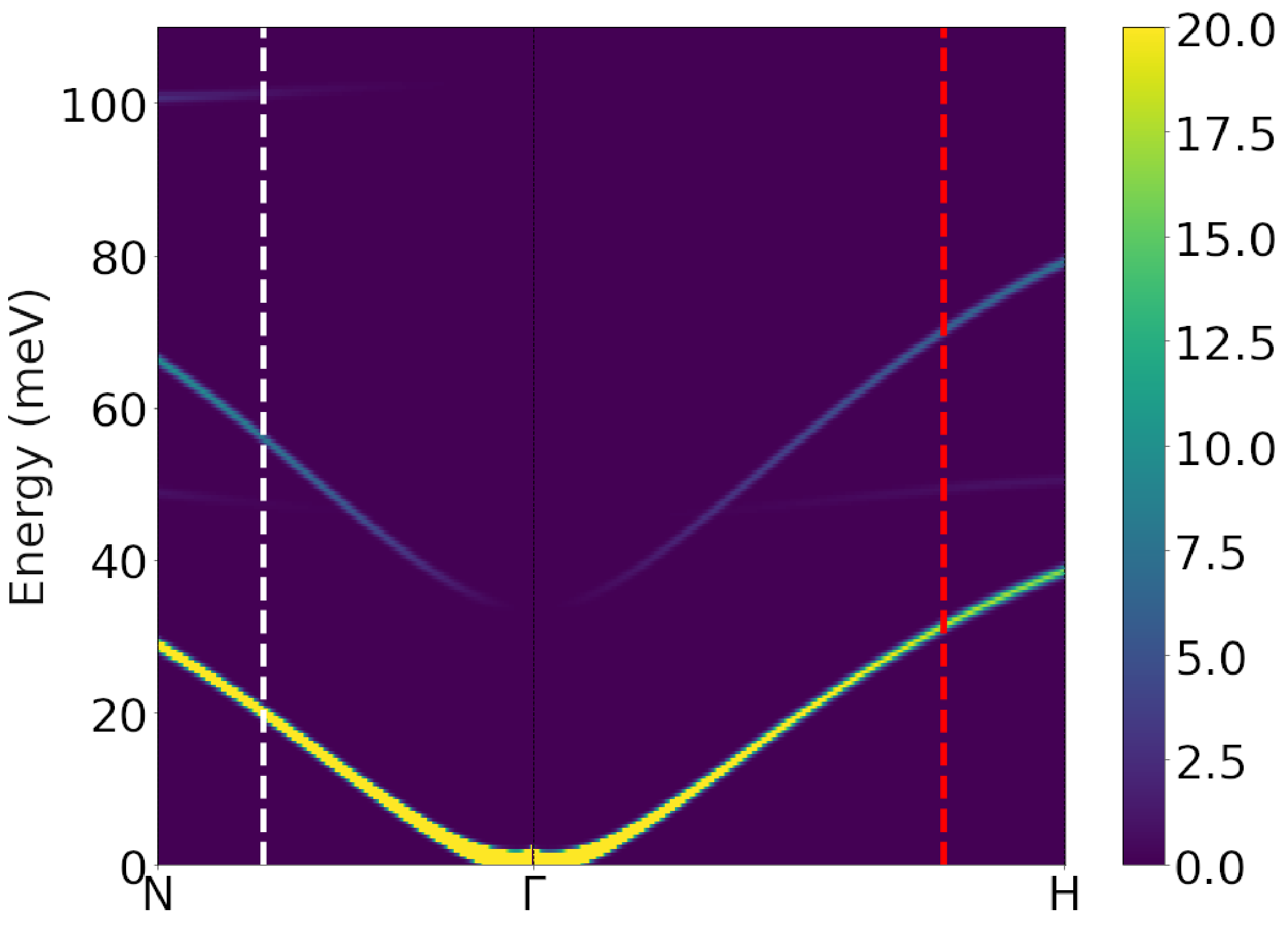}

  \label{fig:NiO_2}
\end{subfigure}\hfil 
\begin{subfigure}{0.333333\textwidth}
    \raggedright
    (c) $\theta = \pi/2$
  \includegraphics[width=\linewidth]{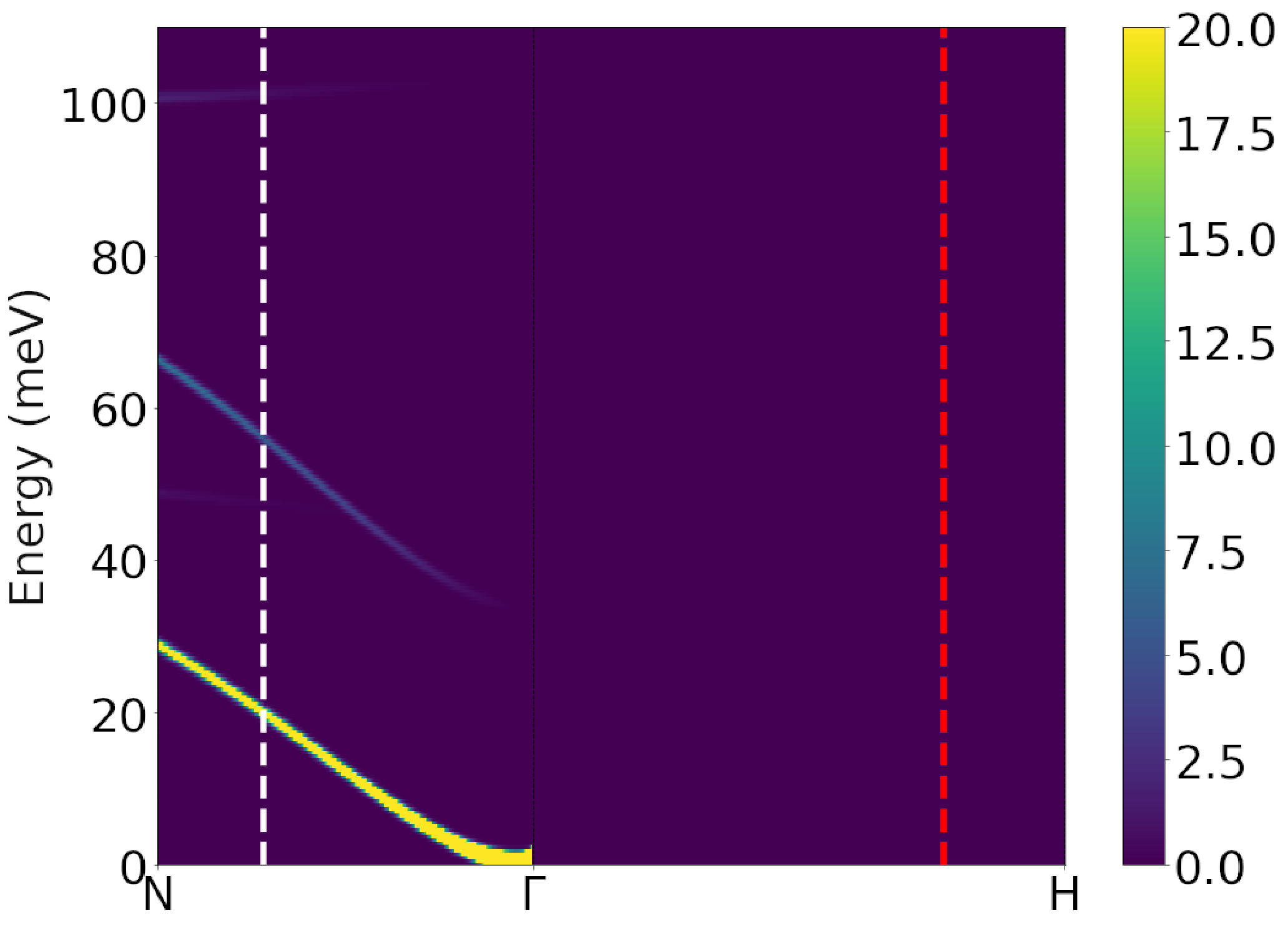}

  \label{fig:NiO_3}
\end{subfigure}

\medskip
\begin{subfigure}{0.3333336\textwidth}
    \raggedright
    (d) $\theta = 7\pi/10$
  \includegraphics[width=\linewidth]{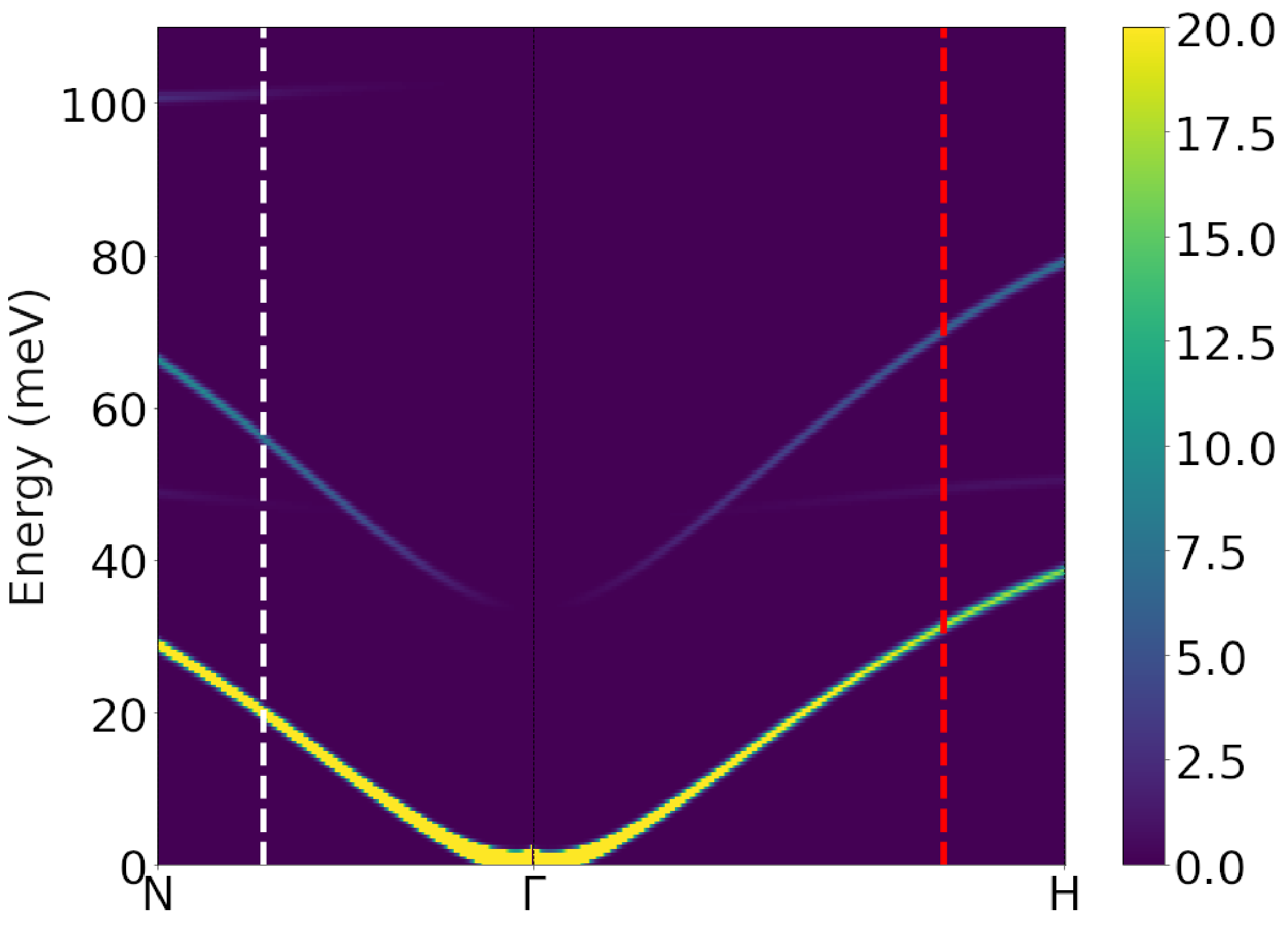}

  \label{fig:NiO_4}
\end{subfigure}\hfil 
\begin{subfigure}{0.333333\textwidth}
    \raggedright
    (e) 
  \includegraphics[width=\linewidth]{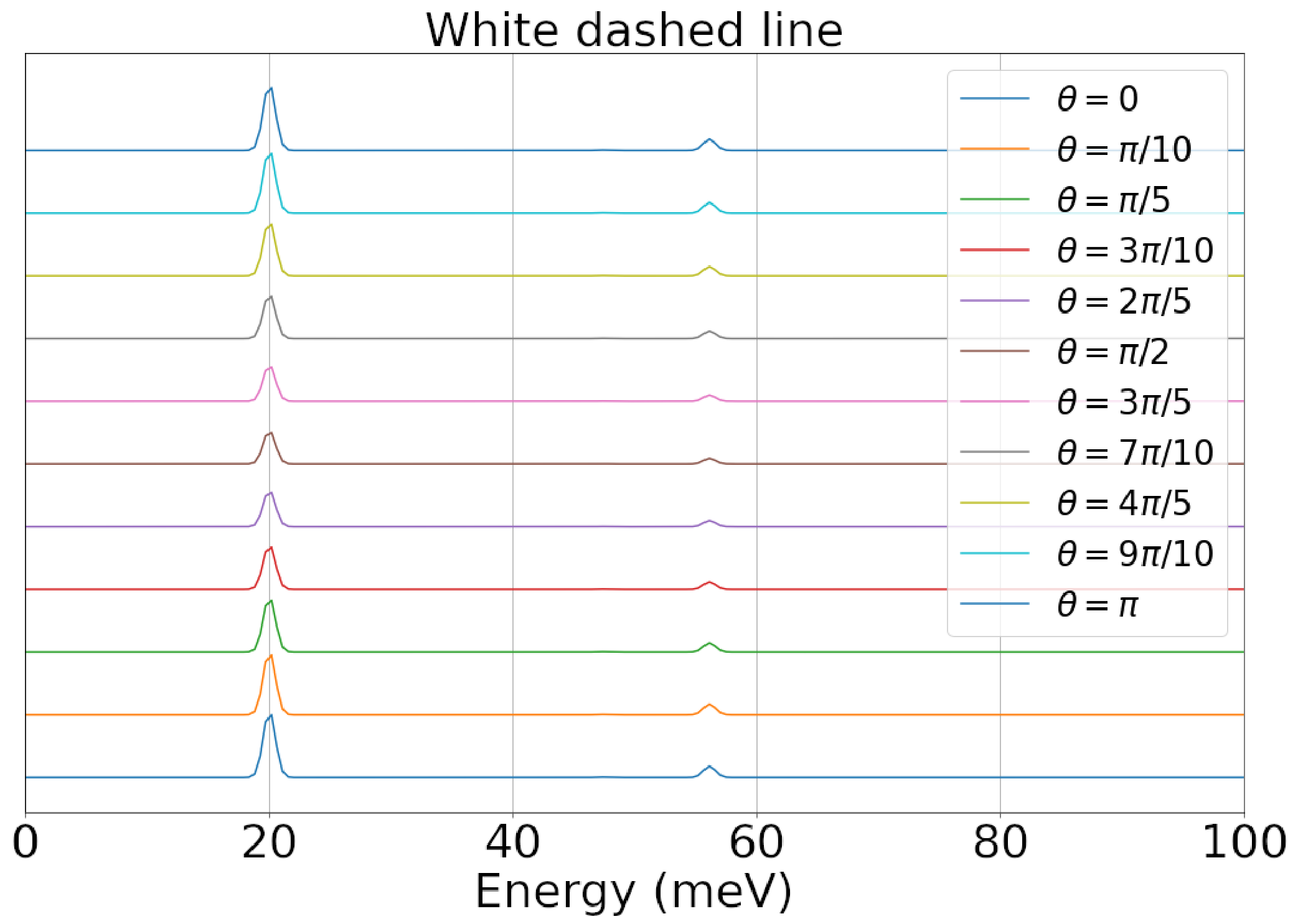}

  \label{fig:NiO_5}
\end{subfigure}\hfil 
\begin{subfigure}{0.333333\textwidth}
    \raggedright
    (f)
  \includegraphics[width=\linewidth]{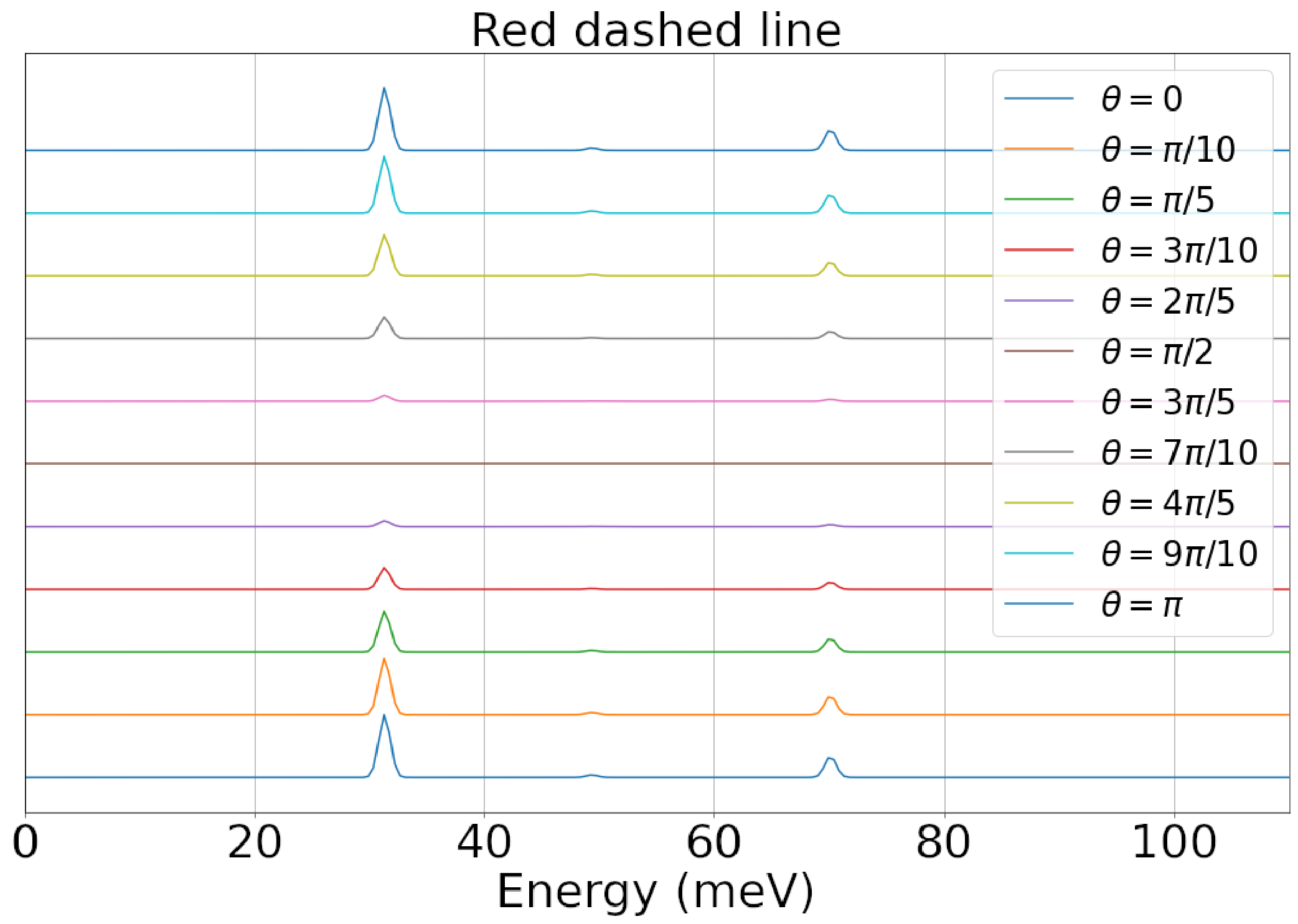}

  \label{fig:NiO_6}
\end{subfigure}
\caption{a-d) Charge-related EELS, for varying relative angles $\theta$ between the probe's wave vector and the Nèel vector. e-f) Angle dependent intensity for a particular point in momentum space, showing a strong angle dependence on the point represented by the red dashed line, but not on the point represented by the white dashed line.}
\label{fig:YIG_orientation}
\end{figure*}

\section{Conclusion}\label{Conclusion}

The proposed methodology is intended to facilitate the distinction of magnon-related pecks in the EELS experiments when paired with the evaluation of the phonon EEL spectrum. The high spatial resolution united with the magnetic moment orientation sensitive nature of the non-spin polarized EELS spectrum, and given the difference in $q$ dependence, a particular choice of spectra taken from different, but related, scattering vectors can be used to probe local differences in the orientation of the Curie/Neel vector relative to the electron beam momentum. The dependence of the double differential cross-section of the charge-based interaction with the electron beam momentum shows that the magnon-related peaks would be more intense for higher acceleration voltages.

\section{Acknowledgement}\label{ack}

This project was undertaken on the Viking Cluster, which is a high-performance computing facility provided by the University of York. We are grateful for computational support from the University of York High-Performance Computing service, Viking and the Research Computing team.



\vspace{2cm}

\bibliographystyle{unsrt}

\bibliography{bibliography}

\begin{thebibliography}{10}

\bibitem{Mahmoud2020}
Abdulqader Mahmoud, Florin Ciubotaru, Frederic Vanderveken, Andrii~V. Chumak,
  Said Hamdioui, Christoph Adelmann, and Sorin Cotofana.
\newblock Introduction to spin wave computing.
\newblock {\em Journal of Applied Physics}, 128(16), October 2020.

\bibitem{Barman_2021}
Anjan Barman, Gianluca Gubbiotti, S~Ladak, A~O Adeyeye, M~Krawczyk, J~Gräfe,
  C~Adelmann, S~Cotofana, A~Naeemi, V~I Vasyuchka, B~Hillebrands, S~A Nikitov,
  H~Yu, D~Grundler, A~V Sadovnikov, A~A Grachev, S~E Sheshukova, J-Y Duquesne,
  M~Marangolo, G~Csaba, W~Porod, V~E Demidov, S~Urazhdin, S~O Demokritov,
  E~Albisetti, D~Petti, R~Bertacco, H~Schultheiss, V~V Kruglyak, V~D Poimanov,
  S~Sahoo, J~Sinha, H~Yang, M~Münzenberg, T~Moriyama, S~Mizukami, P~Landeros,
  R~A Gallardo, G~Carlotti, J-V Kim, R~L Stamps, R~E Camley, B~Rana, Y~Otani,
  W~Yu, T~Yu, G~E~W Bauer, C~Back, G~S Uhrig, O~V Dobrovolskiy, B~Budinska,
  H~Qin, S~van Dijken, A~V Chumak, A~Khitun, D~E Nikonov, I~A Young, B~W
  Zingsem, and M~Winklhofer.
\newblock The 2021 magnonics roadmap.
\newblock {\em Journal of Physics: Condensed Matter}, 33(41):413001, aug 2021.

\bibitem{Kerr2017}
Paul~S. Keatley, Thomas H.~J. Loughran, Euan Hendry, William~L. Barnes,
  Robert~J. Hicken, Jeffrey~R. Childress, and Jordan~A. Katine.
\newblock {A platform for time-resolved scanning Kerr microscopy in the
  near-field}.
\newblock {\em Review of Scientific Instruments}, 88(12):123708, 12 2017.

\bibitem{BLS2015}
Thomas Sebastian, Katrin Schultheiss, Björn Obry, Burkard Hillebrands, and
  Helmut Schultheiss.
\newblock Micro-focused brillouin light scattering: imaging spin waves at the
  nanoscale.
\newblock {\em Frontiers in Physics}, 3, 2015.

\bibitem{Hage2019}
F.~S. Hage, D.~M. Kepaptsoglou, Q.~M. Ramasse, and L.~J. Allen.
\newblock Phonon spectroscopy at atomic resolution.
\newblock {\em Phys. Rev. Lett.}, 122:016103, Jan 2019.

\bibitem{Hage2020}
F.~S. Hage, G.~Radtke, D.~M. Kepaptsoglou, M.~Lazzeri, and Q.~M. Ramasse.
\newblock Single-atom vibrational spectroscopy in the scanning transmission
  electron microscope.
\newblock {\em Science}, 367(6482):1124--1127, 2020.

\bibitem{Hoglund2022-qm}
Eric~R Hoglund, De-Liang Bao, Andrew O'Hara, Sara Makarem, Zachary~T
  Piontkowski, Joseph~R Matson, Ajay~K Yadav, Ryan~C Haislmaier, Roman
  Engel-Herbert, Jon~F Ihlefeld, Jayakanth Ravichandran, Ramamoorthy Ramesh,
  Joshua~D Caldwell, Thomas~E Beechem, John~A Tomko, Jordan~A Hachtel,
  Sokrates~T Pantelides, Patrick~E Hopkins, and James~M Howe.
\newblock Emergent interface vibrational structure of oxide superlattices.
\newblock {\em Nature}, 601(7894):556--561, 2022.

\bibitem{Loudon2012}
J.~C. Loudon.
\newblock Antiferromagnetism in nio observed by transmission electron
  diffraction.
\newblock {\em Phys. Rev. Lett.}, 109:267204, Dec 2012.

\bibitem{Rusz2021}
Keenan Lyon, Anders Bergman, Paul Zeiger, Demie Kepaptsoglou, Quentin~M.
  Ramasse, Juan~Carlos Idrobo, and J\'an Rusz.
\newblock Theory of magnon diffuse scattering in scanning transmission electron
  microscopy.
\newblock {\em Phys. Rev. B}, 104:214418, Dec 2021.

\bibitem{PLOTKINSWING2020}
Benjamin Plotkin-Swing, George~J. Corbin, Sacha {De Carlo}, Niklas Dellby,
  Christoph Hoermann, Matthew~V. Hoffman, Tracy~C. Lovejoy, Chris~E. Meyer,
  Andreas Mittelberger, Radosav Pantelic, Luca Piazza, and Ondrej~L. Krivanek.
\newblock Hybrid pixel direct detector for electron energy loss spectroscopy.
\newblock {\em Ultramicroscopy}, 217:113067, 2020.

\bibitem{KRIVANEK201960}
O.L. Krivanek, N.~Dellby, J.A. Hachtel, J.-C. Idrobo, M.T. Hotz,
  B.~Plotkin-Swing, N.J. Bacon, A.L. Bleloch, G.J. Corbin, M.V. Hoffman, C.E.
  Meyer, and T.C. Lovejoy.
\newblock Progress in ultrahigh energy resolution eels.
\newblock {\em Ultramicroscopy}, 203:60--67, 2019.
\newblock 75th Birthday of Christian Colliex, 85th Birthday of Archie Howie,
  and 75th Birthday of Hannes Lichte / PICO 2019 - Fifth Conference on
  Frontiers of Aberration Corrected Electron Microscopy.

\bibitem{Sturm1993}
K.~Sturm.
\newblock Dynamic structure factor: An introduction.
\newblock {\em Zeitschrift für Naturforschung A}, 48(1-2):233--242, 1993.

\bibitem{Stephen1984}
Stephen~W. Lovesey.
\newblock {\em Theory of Neutron Scattering from Condensed Matter Volume II:
  Polarization Effects and Magnetic Scattering}.
\newblock Clarendon Press, 1984.

\bibitem{Lovesley}
S~W Lovesey.
\newblock {\em Theory of neutron scattering from condensed matter. Vol. 2.
  Polarization effect and magnetic scattering}.
\newblock Jan 1984.

\bibitem{Squires2012}
G.~L. Squires.
\newblock {\em Introduction to the Theory of Thermal Neutron Scattering}.
\newblock Cambridge University Press, March 2012.

\bibitem{Mendis2022}
B.G. Mendis.
\newblock Quantum theory of magnon excitation by high energy electron beams.
\newblock {\em Ultramicroscopy}, 239:113548, September 2022.

\bibitem{Fourier}
Gregory~S. Adkins.
\newblock Three-dimensional fourier transforms, integrals of spherical bessel
  functions, and novel delta function identities, 2013.

\bibitem{Thz_Spectroscopy}
Randy~S Fishman, Jaime~A Fernandez-Baca, and Toomas Rõõm.
\newblock {\em Spin-Wave Theory and its Applications to Neutron Scattering and
  THz Spectroscopy}.
\newblock 2053-2571. Morgan and Claypool Publishers, 2018.

\bibitem{Kubler1988}
J~Kubler, K~H Hock, J~Sticht, and A~R Williams.
\newblock Density functional theory of non-collinear magnetism.
\newblock {\em Journal of Physics F: Metal Physics}, 18(3):469–483, March
  1988.

\bibitem{Nicholls2019}
R.~J. Nicholls, F.~S. Hage, D.~G. McCulloch, Q.~M. Ramasse, K.~Refson, and
  J.~R. Yates.
\newblock Theory of momentum-resolved phonon spectroscopy in the electron
  microscope.
\newblock {\em Phys. Rev. B}, 99:094105, Mar 2019.

\bibitem{HudsonINS}
Bruce~S. Hudson.
\newblock Inelastic neutron scattering: A tool in molecular vibrational
  spectroscopy and a test of ab initio methods.
\newblock {\em The Journal of Physical Chemistry A}, 105(16):3949--3960, 2001.

\bibitem{Princep2017}
Andrew~J. Princep, Russell~A. Ewings, Simon Ward, Sandor Tóth, Carsten Dubs,
  Dharmalingam Prabhakaran, and Andrew~T. Boothroyd.
\newblock The full magnon spectrum of yttrium iron garnet.
\newblock {\em npj Quantum Materials}, 2(1), November 2017.

\bibitem{Brown}
P.J. Brown.
\newblock Magnetic form factors.
\newblock Available at \url{https://www.ill.eu/sites/ccsl/ffacts/}.

\bibitem{NeutronFlux}
Stewart~F. Parker.
\newblock Inelastic neutron scattering, instrumentation*.
\newblock In John~C. Lindon, editor, {\em Encyclopedia of Spectroscopy and
  Spectrometry (Second Edition)}, pages 1035--1044. Academic Press, Oxford,
  second edition edition, 2010.

\bibitem{ElectronFlux}
Martha Ilett, Mark S’ari, Helen Freeman, Zabeada Aslam, Natalia Koniuch,
  Maryam Afzali, James Cattle, Robert Hooley, Teresa Roncal-Herrero, Sean~M.
  Collins, Nicole Hondow, Andy Brown, and Rik Brydson.
\newblock Analysis of complex, beam-sensitive materials by transmission
  electron microscopy and associated techniques.
\newblock {\em Philosophical Transactions of the Royal Society A: Mathematical,
  Physical and Engineering Sciences}, 378(2186):20190601, October 2020.

\end{thebibliography}

\end{document}